\documentclass[a4paper,11pt]{article}
\usepackage{jheppub}

\usepackage{graphicx}
\usepackage{amsmath}
\usepackage{mathrsfs}
\usepackage{xspace}
\usepackage{subcaption}
\usepackage{slashed}
\usepackage{braket}

\newcommand{\us}{\text{us}}

\newcommand{\tvec}[1]{\boldsymbol{#1}}

\newcommand{\csgn}[2]{\eta_{#1}(#2)}

\newcommand{\ms}{\mskip 1.5mu}
\newcommand{\bs}{\mskip -1.5mu}

\newcommand{\pr}[2]{{}^{#1}\bs #2}      
\newcommand{\prb}[2]{{}^{#1}\! #2}      
\newcommand{\prn}[2]{{}^{#1} #2}        

\newcommand{\half}{{\textstyle\frac{1}{2}}}

\usepackage{xcolor}

\usepackage[normalem]{ulem}
\newcommand{\new}[1]{\textcolor{black}{#1}}

\allowdisplaybreaks[2]

\title{Transverse momentum dependence in double parton scattering}
\author[a]{Jonathan R.~Gaunt}
\author[b]{and Tomas Kasemets}

\affiliation[a]{CERN Theory Division, 1211 Geneva 23, Switzerland}
\affiliation[b]{PRISMA Cluster of Excellence \& Mainz Institute for Theoretical Physics Johannes Gutenberg University, 55099 Mainz, Germany}

\emailAdd{jonathan.richard.gaunt@cern.ch}
\emailAdd{kasemets@uni-mainz.de}

\preprint{\vbox{
\hbox{MITP/18-126, CERN-TH-2018-281}}}

\abstract{In this review, we describe the status of transverse momentum dependence (TMD) in double parton scattering (DPS). The different regions of TMD DPS are discussed, and expressions given for the DPS cross section contributions that make use of as much perturbative information as possible. The regions are then combined with each other as well as with single parton scattering to obtain a complete expression for the cross section. Particular emphasis is put on the differences and similarities to transverse momentum dependence in single parton scattering. We further discuss the status of the factorisation proof for double colour singlet production in DPS, which is now on a similar footing to the proofs for TMD factorisation in single Drell-Yan, discuss parton correlations and give an outlook on possible research on DPS in the near future.}

\begin{document}

\maketitle

\section{Introduction}

Double parton scattering (DPS) is the process in which one has two hard scatterings, producing two sets of particles that we can label as `$1$' and `$2$', in an individual proton-proton collision\footnote{\new{One allows any possibility for the final-state particles accompanying `$1$' and `$2$'; these are often denoted by the symbol $X$ and are typically the products of additional soft scatterings and soft/collinear radiation from the partons active in the hard processes. The proof of factorisation for double colour-singlet production in DPS relies on this inclusive definition \cite{Diehl:2015bca}.}}. The region in which the transverse momenta \new{ of systems 1 and 2}, $\tvec{q}_1$ and $\tvec{q}_2$, are small is particularly important in studies of DPS, since DPS is especially prominent in this region compared to the usual single parton scattering (SPS) mechanism \cite{Diehl:2011yj, Blok:2011bu} \new{(note that here we use boldface symbols to denote transverse momentum vectors)}. Indeed, many experimental extractions of DPS use variables sensitive to this `double back-to-back' configuration; for example, the $\Delta_{ij}^{p_T}$ variable in \cite{Aaboud:2016dea}. In the small $\tvec{q}_1, \tvec{q}_2$  region a description in terms of double parton transverse momentum dependent (TMD) distributions is appropriate. There are many parallels between the treatment of TMD cross sections in single parton scattering (SPS) and double parton scattering (DPS). There are however also clear differences, with direct physical consequences. In this review, we aim to highlight these differences and similarities in order to facilitate researchers interested in spin and TMD physics to make important contributions to the field of DPS. 

While TMD factorisation in SPS has been rigorously proven for colour singlet production, see e.g. \cite{Collins:2011zzd}, it runs into problems for hadron collisions producing coloured final states \cite{Rogers:2010dm,Collins:2007nk,Collins:2007jp}. These issues are expected to be important also for DPS, and we will therefore restrict ourselves to double colour-singlet production, i.e. where each of the two hard collisions separately produce a colour singlet final state.

\begin{figure}[t]
\begin{center}
\includegraphics[width=0.40\textwidth]{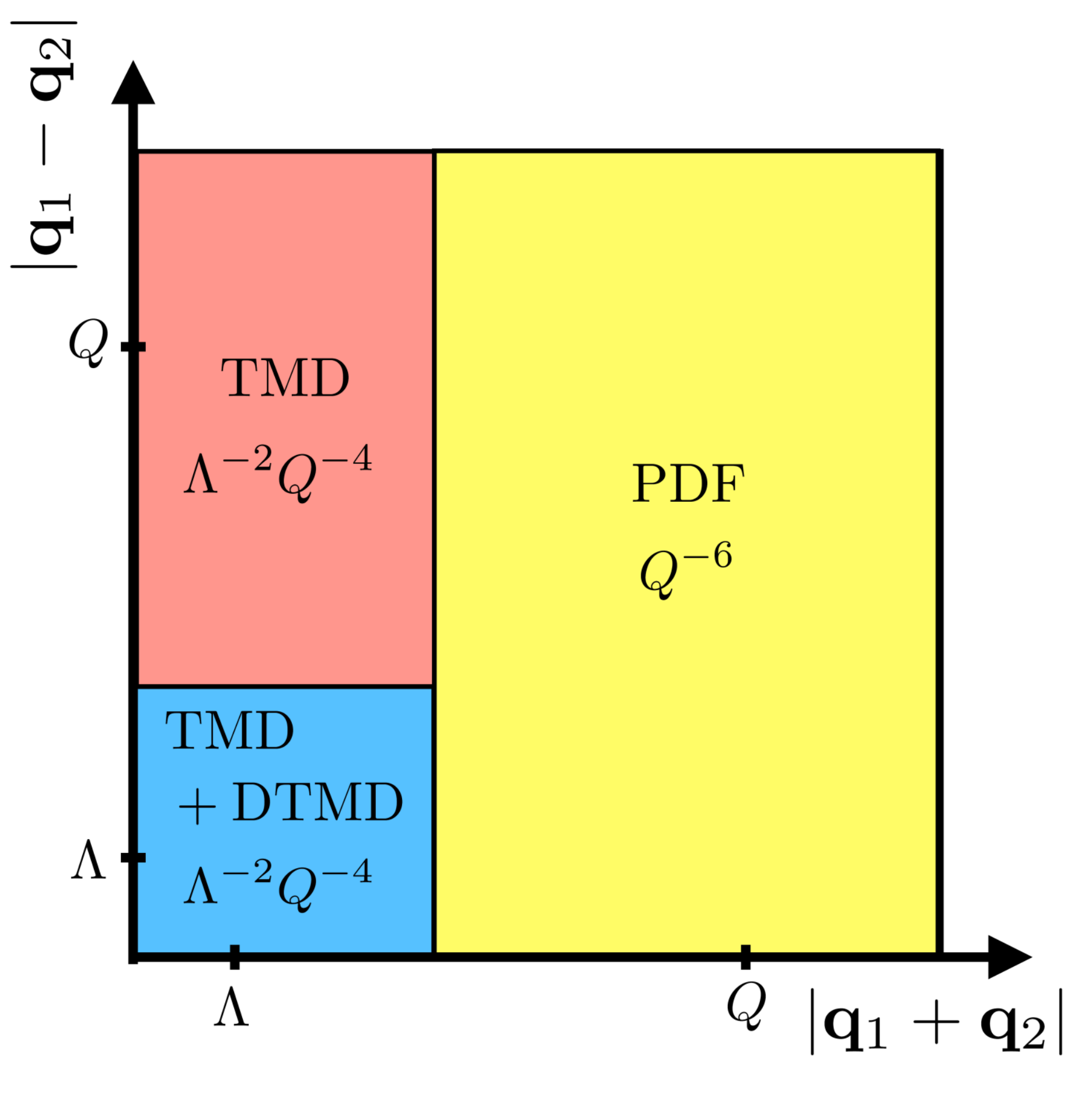}
\caption{\label{fig:qTRegions} Transverse momentum regions and descriptions of the initial state in the leading power factorisation theorems. The power counting behaviour of the differential cross section in each region is indicated on the figure.}
\end{center}
\end{figure}

In the production of two colour singlets, such as two vector bosons, the TMD SPS factorisation theorem can be applied as usual to study the region where the sum of the two transverse momenta is small. If however, the transverse momenta of both bosons are measured to be much smaller than the hard scale, standard TMD factorisation alone is no longer sufficient. For these observables, DPS contributes at the same power as SPS, and no leading-power factorisation theorem can be derived without simultaneously taking care of SPS and DPS, including their interference. An overview of the different factorisation theorems in hadron collisions and the treatment of the initial state in different regions of the sum and difference of the transverse momenta of the two colour singlets is shown in Figure~\ref{fig:qTRegions}. 
Once we integrate over the transverse momentum difference $\tvec{q}_1 - \tvec{q}_2$ DPS is degraded to a power correction to the SPS cross section. Nonetheless, there are several processes in high energy collisions where DPS can compete with or surpass the SPS contribution even for the total cross section. This is usually due to enhancements caused by the large increase in parton densities at small momentum fractions and/or that the SPS cross section is suppressed by additional small coupling constants. 

An important issue to address when describing processes which can receive significant contributions from DPS is the consistent combination of SPS and DPS, avoiding double counting. First approaches to this problem are described in \cite{Blok:2011bu, Blok:2013bpa}, \cite{Gaunt:2011xd, Gaunt:2012dd} and \cite{Manohar:2012pe}, but these approaches suffer from the drawback that the process does not factorise in the usual sense into hard cross sections and parton densities for each hadron. A new approach which maintains the usual factorisation (and has certain other advantages) was put forward in \cite{Diehl:2017kgu}. 

A description of the transverse momentum dependent DPS cross section was pursued at the leading logarithmic (LL) level in \cite{Blok:2011bu}, which uses a generalisation of the Dokshitzer-Diakanov-Troian (DDT) formula \cite{Dokshitzer:1978hw} to DPS, and takes the former approach to handling DPS/SPS double counting. It is known from the SPS case that beyond-LL corrections are in practice important -- for example, the DDT formula anticipates a dip in $d\sigma/d|\tvec{q}^2|$ as \new{the transverse momentum of the final-state colour singlet system} $\tvec{q}$ approaches $0$, which is not observed in practice (see e.g. \cite{Ellis:1991qj}). In \cite{Buffing:2017mqm} a framework was developed that holds beyond LL and uses the approach of \cite{Diehl:2017kgu} for handling DPS/SPS double counting. In this paper the ingredients required for an NLL description of transverse momentum distributions were also computed. In this review we will focus on the latter approach.

We note in passing that double parton distributions depending on transverse momentum arguments, sometimes referred to as `unintegrated' double parton distributions (UDPDFs), appear in approaches designed to describe the DPS cross section at small $x$ (see e.g. \cite{Maciula:2018mig,Kutak:2016ukc,Golec-Biernat:2016vbt,Elias:2017flu}). We will not discuss further such approaches, nor the associated UDPDFs, here.

The TMD distributions in DPS (DTMDs) depends on the longitudinal momentum fractions $x_i$ carried by the interacting partons and two transverse distances $\tvec{z}_i$ which are the DPS analogs of the single TMD `impact parameter' $\tvec{b}$. In addition to these, the distributions depend on the average transverse distance $\tvec{y}$ between the partons. This distance must be equal in the two DTMDs to ensure the two partons are in the same transverse region in each set of colliding partons. In the cross section, the DTMDs are integrated over this common $\tvec{y}$ value.

In the rest of this review, we will take a closer look at the current status of TMD and spin physics in DPS. In section~\ref{sec:fthrm} we give the factorisation theorem for the TMD DPS cross section, discuss the different ingredients and their scale evolution. The different regions of TMD DPS, matching calculations in these regions and the combination of regions including both DPS and SPS are discussed in section~\ref{sec:regions}. The status of factorisation proofs for DTMDs is reviewed in section~\ref{sec:factstat}. We discuss interparton correlations in section~\ref{sec:corr}, before outlining a few promising future research directions in section~\ref{sec:for}.

\section{Factorisation theorem and evolution equations} \label{sec:fthrm}
The TMD DPS cross section formula can be written in terms of two hard coefficients and two DTMDs as \cite{Buffing:2017mqm}:
\begin{align}
	\label{TMD-Xsect}
& \frac{d\sigma_{\text{DPS}}}{dx_1\, dx_2\, d\bar{x}_1\, d\bar{x}_2\,
          d^2\tvec{q}_1\, d^2\tvec{q}_2} = \frac{1}{C}\,
        \sum_{a_1,a_2,b_1,b_2} \!\!\!  \hat{\sigma}_{a_1 b_1}(Q_1,
        \mu_1)\, \hat{\sigma}_{a_2 b_2}(Q_2, \mu_2)\,
\nonumber \\ 
&       \qquad \times \int \frac{d^2\tvec{z}_1}{(2\pi)^2}\,
        \frac{d^2\tvec{z}_2}{(2\pi)^2}\, d^2\tvec{y}\; e^{-i\tvec{q}_1
          \tvec{z}_1 -i\tvec{q}_2 \tvec{z}_2}\, W_{a_1 a_2 b_1
          b_2}(\bar{x}_i,x_i, \tvec{z}_i,\tvec{y}; \mu_i,\nu) \,,
\end{align}
with
\begin{align}
	\label{eq:W-def}
& W_{a_1 a_2 b_1 b_2}(\bar{x}_i,x_i,\tvec{z}_i,\tvec{y};\mu_i,\nu)
 = \Phi(\nu \tvec{y}_+)\, \Phi(\nu \tvec{y}_-)
\nonumber \\[0.2em]
 &\qquad \times \sum_{R} \csgn{a_1 a_2}{R}\; \pr{R}{ F_{b_1
            b_2}(\bar{x}_i,\tvec{z}_i,\tvec{y};\mu_i,\bar{\zeta})}\,
        \pr{R}{F_{a_1 a_2}(x_i,\tvec{z}_i,\tvec{y};\mu_i,\zeta)} \,.
\end{align}
We
use light-cone coordinates $w^\pm = (w^0 \pm w^3) /\sqrt{2}$ and the
transverse component $\tvec{w} = (w^1, w^2)$ for any four-vector $w$. \new{For the production of two electroweak gauge bosons $pp \to V_1 + V_2 + X$ with $V_i = \gamma^*, Z, W$, $Q_i^2$ is the squared invariant mass of $V_i$, and $x_i, \bar{x}_i$ are related to the $Q_i^2$ and the rapidities of the produced vector bosons $Y_i$ as follows:}
\begin{align}
x_i = \sqrt{\dfrac{Q_i^2}{s}}e^{Y_i} \qquad \bar{x}_i = \sqrt{\dfrac{Q_i^2}{s}}e^{-Y_i}
\end{align}

Although the structure of this formula is very similar to the TMD factorisation for SPS, there are several interesting differences as we will see as we have a closer look at the different ingredients. $\pr{R}{F_{a_1 a_2}(x_i,\tvec{z}_i,\tvec{y};\mu_i,\zeta)}$ is the TMD for partons $a_1$ and $a_2$ to be found inside the proton, with longitudinal momentum fractions $x_1$  and $x_2$ respectively and at transverse positions $\tvec{y}\pm\tvec{z}_1/2$ and $\pm \tvec{z}_2/2$ (where the plus sign corresponds to the position in the amplitude, and the minus sign corresponds to the position in the conjugate). $R$ labels the colour representation of a a parton in the amplitude coupled to its partner in the conjugate amplitude. The motivation for this choice of colour decomposition (instead of coupling the two partons in the amplitude) is the separation of the colour singlet $R=1$ as the representation that is free from any colour correlations between the two hard interactions. The DTMD depends on three scales, two separate renormalization scales $\mu_1$ and $\mu_2$, and a rapidity scale $\zeta$, as will be shown in more detail when we discuss their definitions and evolution equations. The two functions $\Phi(\nu \tvec{y}_\pm)$, with
\begin{align}
	\label{y-pm-def}
\tvec{y}_{\pm} = \tvec{y} \pm \half (\tvec{z}_1 - \tvec{z}_2) \,
\end{align}
regulates the UV-region where SPS and DPS overlap as will be discussed further when we return to the consistent combination of SPS and DPS in section~\ref{sec:SPScombi}. 
The factor
$\csgn{a_1 a_2}{R}$ is equal to $1$ except for very specific combinations of parton flavours and colour representations (see \cite{Buffing:2017mqm}). The subprocess cross sections are denoted by $\hat\sigma_{a_i b_i}(Q_i^2, \mu_i^2)$.
The sum over $a_1,a_2,b_1,b_2$ in \eqref{TMD-Xsect} runs over both parton
species and polarisations. $C$ is a combinatorial factor equal to 2 if the final states of the two hard processes are indistinguishable and 1 otherwise.

\subsection{Definitions of double parton distributions}
In TMD measurements in SPS, it is well known that the collinear and soft momentum regions individually contain rapidity divergences. To tame these divergences and obtain one function describing each of the hadrons, the soft function is split up and combined with the collinear regions. There are several techniques for this procedure depending on the choice of regulator used for the rapidity divergences, but in essence it boils down to separating soft radiation on each side of a rapidity parameter \cite{Collins:2011zzd,Echevarria:2016scs}. This procedure generates another scale in which the TMDs evolve. The story for DTMDs is in many aspects equivalent, with a number of complications due to the more complicated colour structure leading in particular to soft factors which are matrices in colour space. 
For two partons $a_1$ and $a_2$, the unsubtracted (i.e. before cancellations of rapidity divergences) DTMDs for a right moving proton ($p^3>0$) are defined in
terms of matrix elements as \cite{Diehl:2011yj,Diehl:2011tt}
\begin{align}
\label{eq:dpds}
	F_{\us, a_1a_2}(x_1,x_2,\tvec{z}_1,\tvec{z}_2,\tvec{y}) & = 2p^+
        (x_1\ms p^+)^{-n_1}\, (x_2\ms p^+)^{-n_2} \int
        \frac{dz^-_1}{2\pi}\, \frac{dz^-_2}{2\pi}\, dy^-\,
          e^{i\ms ( x_1^{} z_1^- + x_2^{} z_2^-)\ms p^+}
\nonumber \\
 & \quad \times
    \langle\ms p \ms|\, \mathcal{O}_{a_1}(y,z_1)\, \mathcal{O}_{a_2}(0,z_2)
    \,|\ms p \ms\rangle \,,
\end{align}
where $n_i = 1$ if parton number $i$ is a gluon and $n_i = 0$ otherwise.  It is understood that
$\tvec{p} = \tvec{0}$ and that the proton polarisation is
averaged over. 
The operators for quarks (see e.g. \cite{Diehl:2011yj} for the gluon equivalent) in a right moving proton read
\begin{align}
\label{eq:quark-ops}
\mathcal{O}_{a}(y,z) &=
  \bar{q}\bigl( y - \half z \bigr)\, W^\dagger \bigl(y-\half z, v_L \bigr) \,
  \Gamma_{a} \, W \bigl(y+\half z, v_L \bigr) \, q\bigl( y + \half z \bigr)
\Big|_{z^+ = y^+_{} = 0}
\end{align}
with spin projections $\Gamma_q  = \half \gamma^+$ for an unpolarised quark. 
The field with
argument $y + \half z$ in $\mathcal{O}_{q}(y,z)$ is associated with a
quark in the amplitude of a double scattering process and the field with
argument $y - \half z$ with a quark in the complex conjugate amplitude. $W(y, v)$ is a past-pointing Wilson line in the direction $v$ originating at the point $y$. The vector $v_L$ is a spacelike, purely longitudinal vector that is nearly aligned with the minus light-cone direction; this vector is tilted slightly off the lightcone to regulate the rapidity divergences. \new{For the left-moving proton one uses a vector $v_R$ nearly aligned with the plus light-cone direction.}


In processes producing colourless particles, one needs
the soft factor 
\begin{align}
	\label{soft-def}
&S_{qq}(\tvec{z}_1,\tvec{z}_2,\tvec{y}; v_L, v_R) 
\nonumber \\[0.2em]
& \qquad = \bigl\langle\, 0 \,\big|\ms
  O_{S,q}(\tvec{y},\tvec{z}_1; v_L,v_R) \,
  O_{S,q}(\tvec{0},\tvec{z}_2; v_L,v_R) \,
  \big|\, 0 \,\bigr\rangle
\end{align}
with
\begin{align}
  \label{WL-ops}
& O_{S,q}(\tvec{y},\tvec{z}; v_L,v_R)
\nonumber \\
& \qquad
   = \bigl[ W^{}(\tvec{y}+\half \tvec{z},v_L)\,
     W^\dagger(\tvec{y}+\half \tvec{z},v_R) \bigr] \,
     \bigl[ W^{}(\tvec{y}-\half \tvec{z},v_R)\,
     W^\dagger(\tvec{y}-\half \tvec{z},v_L) \bigr] \,.
\end{align}
The soft matrix $S$ \new{in \eqref{soft-def}} depends on $v_L$ and $v_R$ only via the difference of the Wilson line rapidities $Y \equiv Y_R - Y_L$\new{: that is, $S(v_L, v_R) = S(Y)$}. \new{The matrix $S(Y)$ can be divided into two pieces at some rapidity $Y_0$ according to \cite{Vladimirov:2017ksc, Buffing:2017mqm}}
\begin{figure}[t]
\begin{center}
\includegraphics[width=1.0\textwidth]{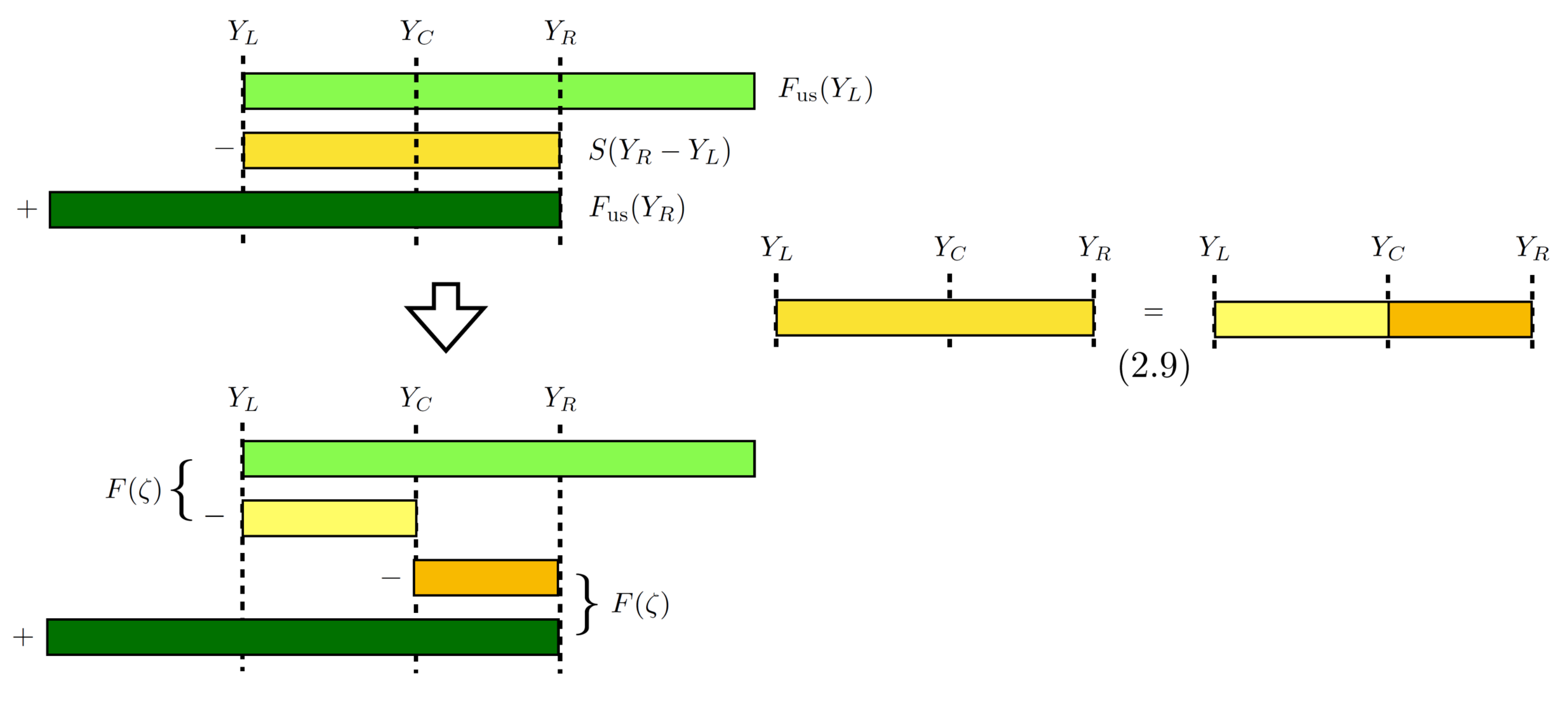}
\caption{\label{fig:rapsub} Rapidity subtractions and definition of right and left moving DTMDs.}
\end{center}
\end{figure}
%
%
%
\begin{align}
	\label{S-decomp}
S(Y) &= s(Y - Y_0)\, s^\dagger(Y_0) & & \text{for $Y \gg 1$ and arbitrary
  $Y_0$} \,.
\end{align}
We have kept the colour indices implicit in these equations, where the colour indices of the soft factor can be projected into a colour matrix in $RR'$, and the unsubtracted DTMDs are  vectors in this colour space.

The subtracted DTMDs, i.e. with the rapidity divergences subtracted through the combination with the soft factor as illustrated in Figure~\ref{fig:rapsub}, are defined as \cite{Buffing:2017mqm}
\begin{align}
  \label{sub-unsub}
\pr{R}{F_{a_1 a_2}}(\zeta)
&= \lim_{Y_L\to -\infty} \sum_{R'}
   \pr{RR'}{ s_{a_1 a_2}^{-1}(Y_C-Y_L) }\;
     \prb{R'}{F_{\us, a_1 a_2}^{}}(Y_L)\,,
\end{align}
for the distributions in a right-moving proton.  $Y_C$ is a central rapidity \new{(typically chosen to be close to zero)}, used to separate left- from right-moving soft radiation, $Y_L \ll Y_C \ll Y_R$. The rapidity scales $\zeta$ and $\bar{\zeta}$ in \eqref{eq:W-def} are related to this central rapidity parameter through
\begin{align}
	\label{zeta-def}
\zeta &= 2 x_1 x_2\, (p^+)^2\, e^{-2 Y_C}\,, \\ 
\bar{\zeta} &= 2 \bar{x}_1 \bar{x}_2\, (\bar{p}^-)^2\, e^{2 Y_C}\,.
\end{align}
where $p^+$ ($\bar{p}^-$) is the plus (minus) momentum of the right- (left-)moving proton.

It can be useful to compare this result to the definition of the subtracted (single parton scattering) TMD, see e.g. chapter 13 in \cite{Collins:2011zzd} and section~3.4 of \cite{Buffing:2017mqm}.
\begin{align}
	\label{eq:unsub_single}
f_{a}(\zeta) &= \lim_{Y_L\to -\infty} s_{a}^{-1}(Y_C - Y_L)\, f_{\us,a}(Y_L) \,,
\end{align}
where the colour matrix $s$ is now reduced to a single function and the product $x_1 x_2$ in $\zeta$ has been replaced by the square of the single parton momentum fraction, $x^2$. 
%

\subsection{RGE evolution and resummation}

The DTMD evolves in two renormalisation scales, related to the two partons, and one rapidity scale  \cite{Buffing:2017mqm}. The renormalisation group
equation for the DTMDs reads
\begin{align}
   \label{RG-TMD-again}
\frac{\partial}{\partial \log\mu_1}\,
  \pr{R}{F_{a_1 a_2}(x_i,\tvec{z}_i,\tvec{y};\mu_i,\zeta)}
&= \gamma_{F, a_1}(\mu_1, x_1\zeta/x_2)\, \pr{R}{F_{a_1
      a_2}(x_i,\tvec{z}_i,\tvec{y};\mu_i,\zeta)}
\end{align}
for the scale $\mu_1$, and in analogy for $\mu_2$. The rapidity scale evolution (Collins-Soper equation) is given by 
\begin{align}
	\label{CS-TMD}
\frac{\partial}{\partial\log \zeta} \pr{R}{F_{a_1
    a_2}}(x_i,\tvec{z}_i,\tvec{y}; \mu_i,\zeta) &= \frac{1}{2} \sum_{R'}
\prb{RR'}{K_{a_1 a_2}(\tvec{z}_i,\tvec{y}; \mu_i)}\, \prb{R'}{F_{a_1
    a_2}(x_i,\tvec{z}_i,\tvec{y}; \mu_i,\zeta)}  \,.
\end{align}
The one-loop results for these kernels are all available in \cite{Buffing:2017mqm}.
The scale dependence of the Collins-Soper kernel is given by the cusp anomalous dimension
\begin{align}
	\label{CS-TMD-RG}
\frac{\partial}{\partial \log\mu_1}\, \prb{RR'}{K_{a_1
    a_2}(\tvec{z}_i,\tvec{y}; \mu_i)} &= - \gamma_{K, a_1}(\mu_1)\,
\delta_{RR'}
\end{align}
and correspondingly for $\mu_2$.  All UV divergences and dependence on the renormalisation scales are contained in the diagonal elements of 
$\prb{RR'}{K}$.  The kernel $K_{a_1 a_2}$
and the anomalous dimensions $\gamma_{F,a}$ and $\gamma_{K,a}$ depend on the
colour representation of the parton (quarks or antiquarks vs.\ gluons) but not
on their flavour or polarisation.
The solution to the rapidity and renormalisation evolution equations, relating the  DTMDs at different scales, reads
\begin{align}
  \label{DTMD-evolved}
  & \pr{R}{F}_{a_1 a_2}(x_i,\tvec{z}_i,\tvec{y};\mu_i,\zeta)
\nonumber \\
  & \quad = \exp\,\biggl\{ \int_{\mu_{01}}^{\mu_1} \frac{d\mu}{\mu}\,
        \biggl[ \gamma_{a_1}(\mu) - \gamma_{K,a_1}(\mu)\,
          \log\frac{\sqrt{x_1\zeta/x_2}}{\mu} \biggr] +
        \pr{1}{K}_{a_1}(\tvec{z}_1;\mu_{01})
    \log\frac{\sqrt{\zeta}}{\sqrt{\zeta_0}}
\nonumber \\
  & \qquad \hspace{1.5em} + \int_{\mu_{02}}^{\mu_2} \frac{d\mu}{\mu}\,
        \biggl[ \gamma_{a_2}(\mu) - \gamma_{K,a_2}(\mu)\,
          \log\frac{\sqrt{x_2\ms \zeta/x_1}}{\mu} \biggr] +
        \pr{1}{K}_{a_2}(\tvec{z}_2;\mu_{02})
    \log\frac{\sqrt{\zeta}}{\sqrt{\zeta_0}} \,\biggr\}
\nonumber \\
  & \qquad \times \sum_{R'} \prb{RR'}{\exp}\,\biggl[ M_{a_1
            a_2}(\tvec{z}_i,\tvec{y})
          \log\frac{\sqrt{\zeta}}{\sqrt{\zeta_0}} \,\biggr]\,
        \prb{R'}{F}_{a_1
          a_2}(x_i,\tvec{z}_i,\tvec{y};\mu_{01},\mu_{02},\zeta_0)\,,
\end{align}
The exponential in the second and third lines is the generalisation to two
partons of the evolution factor for a single parton TMD.  It resums both double and single logarithms.  The last
line in \eqref{DTMD-evolved} describes the mixing between different
colour representations $R$ under rapidity evolution and involves a single
logarithm.  The double logarithms in the evolution of
$\pr{R}{F}_{a_1 a_2}(x_i,\tvec{z}_i,\tvec{y};\mu_i,\zeta)$ are thus the same
as those for a product of two single TMDs.

\section{Regions and matchings}\label{sec:regions}
\subsection{Regions}
\begin{table}
\begin{center} \renewcommand{\arraystretch}{1.2}
\begin{tabular}{cc} \hline
  region  & approximations \\ \hline
DPS, large $\tvec{y}$  & $|\tvec{z}_i| \ll |\tvec{y}|, 1/\Lambda$ \\
DPS, small $\tvec{y}$  & $|\tvec{z}_i|,|\tvec{y}| \ll 1/\Lambda$ \\
SPS &
  $|\tvec{y}_+|, |\tvec{y}_-| \ll |\tvec{z}_i| \ll 1/\Lambda$ \\ \hline
\end{tabular}
\caption{\label{tab:y-regions} Regions of $\tvec{y}$ discussed in the
text. }
\end{center}
\end{table}

Similar to the small impact parameter expansion possible in TMD SPS, the small $\tvec{z}_i$ region of TMD DPS enable additional perturbative calculations and matchings. This is  relevant when the transverse momenta of the vector bosons obey $\Lambda << q_T << Q$, and Fourier oscillations suppresses contributions to the cross section from the non-perturbative region of $z_i$\footnote{Care must be taken if 
$|\tvec{q}_1 \pm \tvec{q}_2|$ is of order $\Lambda$, as this can spoil the cancellations from the Fourier oscillations, see e.g. section~6.1 of \cite{Buffing:2017mqm}.}. The presence of, in particular, the distance scale $\tvec{y}$ leads to different regions of DPS which all contribute to the DTMD cross section. In order to make maximal use of the predictive power of perturbation theory, we will consider three different regions as summarised in Table~\ref{tab:y-regions}. The first is the large-$\tvec{y}$ region, where the distance within the pairs of partons inside the protons is of hadronic size. The second region is for small-$\tvec{y}$ where the distance is of the same order as $\tvec{z}_i$, i.e. $1/q_T$ and finally there is the SPS region where $\tvec{y}$ is of the order of the inverse of the hard scale.

\subsection{Large-$\tvec{y}$}
The large-$\tvec{y}$ region is perhaps the most natural region for DPS. When a perturbative scale inside the DTMDs is provided by $\tvec{z}_i$, the large distance between the two partons allows for separate matching calculations for each of the two partons. This connects the DTMDs with the collinear DPDFs as
\begin{align}
	\label{full-match} 
\pr{R}{F_{a_1 a_2}(x_i, \tvec{z}_i, \tvec{y};\mu_i,\zeta)}
 &= \sum_{b_1, b_2}
    \prn{R}{C_{a_1 b_1}(x_1\!\bs\smash{'},
      \tvec{z}_1;\mu_1, x_1\zeta/x_2)}
\nonumber \\
 & \qquad\quad \underset{x_1}{\otimes}
    \prn{R}{C_{a_2 b_2}(x_2\!\bs\smash{'},
      \tvec{z}_2;\mu_2, x_2\zeta/x_1)} \underset{x_2}{\otimes}
    \pr{R}{F_{b_1 b_2}(x_i\!\bs\smash{'}, \tvec{y}; \mu_i,\zeta)} \,,
\end{align}
with $ \underset{x}{\otimes}$ denoting a convolution in momentum fraction $x$.
This matching is very similar to that of the standard TMD matching onto PDFs, with the same matching coefficients $\prn{R}{C_{a_2 b_2}}$ for the colour singlet contribution, $R=1$. \new{To avoid large logarithms in the coefficients ${^{R}}C_{a_1 b_1}$ this matching should be conducted with $\mu_i \sim \sqrt{\zeta} \sim b_0/|\tvec{z}_i|$; the DTMDs can then be evolved to other scales using \eqref{DTMD-evolved}, whose form simplifies in the $|\tvec{z}_1| ~ |\tvec{z}_2| \ll |y|$ limit to:}
\begin{align}
  \label{small-z-start}
 & \pr{R}{F_{a_1 a_2}(x_i,\tvec{z}_i,\tvec{y};\mu_i,\zeta)}
\nonumber \\
 & \quad = \exp\,\biggl\{ \int_{\mu_{01}}^{\mu_1}
\frac{d\mu}{\mu}\, \biggl[ \gamma_{a_1}(\mu) - \gamma_{K,a_1}(\mu)
\log\frac{\sqrt{x_1\zeta/x_2}}{\mu} \biggr] 
\nonumber \\
 & \hspace{3.55em} + \int_{\mu_{02}}^{\mu_2} \frac{d\mu}{\mu}\,
\biggl[ \gamma_{a_2}(\mu) - \gamma_{K,a_2}(\mu)
\log\frac{\sqrt{x_2\zeta/x_1}}{\mu} \biggr] 
\nonumber \\
 & \hspace{3.55em} + \Bigl[ \pr{R}{K}_{a_1}(\tvec{z}_1;\mu_{01}) +
\pr{R}{K}_{a_2}(\tvec{z}_2;\mu_{02}) + \prb{R}{J}(\tvec{y};\mu_{0i}) \Bigr]
\log\frac{\sqrt{\zeta}}{\sqrt{\zeta_0}} \biggr\} \,
 \pr{R}{F_{a_1 a_2}(x_i,\tvec{z}_i,\tvec{y};\mu_{0i},\zeta_0)}  \,,
\end{align}
where, in particular, the different colour channels no longer mix with each other. The hadronic distance between the two partons does not allow the perturbative evolution to connect the two partons, and therefore can not change their colour representation. The matching in this region can in principle be extrapolated into the region of non-perturbative $z_i$, by extending methods such as the $b^*$ method from TMDs in single parton scattering \cite{Collins:1981va,Collins:1984kg}, although this has not been extensively explored.

\subsection{Small-$\tvec{y}$}
When $\tvec{y}$ becomes small, i.e. $y \sim z_i \sim 1/q_T$, one no longer has a separate matching for each of the two partons.
Following the discussion in \cite{Diehl:2017kgu}, we write
\begin{align}
  \label{DPD-split-intr}
  \pr{R}{F} &= \pr{R}{F}_{\text{spl}} +
      \pr{R}{F}_{\text{int}} \,,
\end{align}
where the short-distance expansion of the terms on the r.h.s.\ involves
proton matrix elements of operators with twist two and twist
four. 

The splitting contribution $F_{\text{spl}}$ describes the case where a
single parton splits into partons $a_1$ and $a_2$
\begin{align}
  \label{split-TMD-coll}
\pr{R}{F}_{a_1 a_2,\, \text{spl}}(x_i,\tvec{z}_i,\tvec{y}; \mu,\mu,\zeta)
&= \frac{\tvec{y}_{+}^{l}\ms \tvec{y}_{-}^{l'}}{ \tvec{y}_{+}^{2}\ms
  \tvec{y}_{-}^{2}}\, \frac{\alpha_s(\mu)}{2\pi^2}\;
\prn{R}{T}_{a_0\to a_1 a_2}^{ll'}
\biggl( \frac{x_1}{x_1+x_2} \biggr)\,
\frac{f_{a_0}(x_1+x_2; \mu)}{x_1+x_2} + \mathcal{O}(\alpha_s^2) \,,
\end{align}
with the one loop coefficients given in \cite{Diehl:2014vaa, Buffing:2017mqm}. A $\zeta$ dependence appears only at order $\alpha_s^2$. 

The term $F_{\text{int}}$ in \eqref{DPD-split-intr} is referred to as the
``intrinsic'' contribution to the DPD and may be thought of as describing
parton pairs $a_1$, $a_2$ in the ``intrinsic'' proton wave function. Unlike $F_{\text{spl}}$, it starts at order $\alpha_s^0$ and reads
\begin{align}
  \label{intr-TMD-tw4}
  \pr{R}{F}_{a_1 a_2,\, \text{int}}(x_i,\tvec{z}_i,\tvec{y};
  \mu,\mu,\zeta) &= \prn{R}{G}_{a_1 a_2}(x_1,x_2,x_2,x_1;\mu)
  + \mathcal{O}(\alpha_s) \,,
\end{align}
where $\prn{R}{G}$ denotes a collinear twist-four distribution.  
Since the corresponding expansion for the intrinsic part of the DPDFs is the same up to $\mathcal{O}(\alpha_s^0)$, one may replace the twist-four functions by the intrinsic part of the DPDFs at leading-order accuracy:
\begin{align}
 \pr{R}{F}_{\text{int},\, a_1 a_2}(x_i,\tvec{z}_i,\tvec{y};
\mu,\mu,\zeta) &= \pr{R}{F}_{\text{int},\, a_1 a_2}(x_i,\tvec{y};
\mu,\mu,\zeta) + \mathcal{O}(\alpha_s) \,,
\end{align}

\new{To avoid large logarithms in the matching coefficients, one should perform these matchings at the scale $\mu_i \sim \sqrt{\zeta} \sim b_0/|\tvec{z}_i| \sim b_0/|\tvec{y}|$; the DTMDs can then be evolved to other scales using \eqref{DTMD-evolved}.}

\subsection{Combination of DPS regions}
The full DPS cross section can be obtained from the combination of the large- and small-$\tvec{y}$ regions through
\begin{align}
  \label{Xsect-DPS}
W_{\text{DPS}}(\nu) &= W_{\text{large $y$}}(\nu') -
  W_{\text{sub}}(\nu') + W_{\text{small $y$}}(\nu) \,,
\end{align}
where we make explicit the choice of cutoff parameters for the $\tvec{y}$
integration, taking $\nu' \sim q_T$ and $\nu \sim Q$. The $W$ terms in the different regions are obtained by replacing the DTMDs in \eqref{eq:W-def} by their approximations in the corresponding regions. The double counting subtraction term is defined as
\begin{align}
  \label{TMD-subtr-term}
W_{\text{sub}} &= W_{\text{small $y$}} \,
\bigl|_{\text{approx.~for}~|\tvec{z}_i| \ll |\tvec{y}|}
\end{align}
with the small-$\tvec{y}$ expression for $W$.  The limit $|\tvec{z}_1|, |\tvec{z}_2| \ll |\tvec{y}|$ should be taken in all
parts of the expression.

In the region
$|\tvec{y}| \gg |\tvec{z}_1|, |\tvec{z}_2|$ the last two terms of \eqref{Xsect-DPS} cancel by
virtue of \eqref{TMD-subtr-term}, and one is left with the first term, which
is designed to give a correct approximation of the cross section there. For
$|\tvec{y}| \sim |\tvec{z}_1|, |\tvec{z}_2|$, the first and second terms
cancel, and the third term gives a correct
approximation of the cross section.  In this way, $W_{\text{DPS}}$ leads to a
correct approximation of the DPS cross section for $|\tvec{y}|$ of order $1/q_T$
and larger.

\subsection{Combination with SPS}\label{sec:SPScombi}
The DPS cross section discussed in the last section can be combined with SPS in a consistent manner following \cite{Diehl:2017kgu}. In essence, the functions $\Phi$ of \eqref{eq:W-def} cut off the DPS cross section where it enters the SPS region. In order to avoid double counting in the DPS region, the part of DPS which is already included in the SPS contributions (i.e. part of the splitting) must be subtracted. 
However, this is not the end of the story, as one has to consistently include the interference between double and single parton scattering (which we shall denote as $\sigma_{\text{DPS/SPS}}$ or $\sigma_{\text{SPS/DPS}}$, depending on which process is in the amplitude/conjugate). This leads to the master formula

\begin{align}
  \label{eq:tot-x-sec}
\sigma
  & = \sigma_{\text{DPS}\rule{0pt}{1.3ex}}
    + \left[\ms \sigma_{\text{DPS/SPS}}
    - \sigma_{\text{DPS},\, y_-\to 0 \rule{0pt}{1.3ex}}
    + \sigma_{\text{SPS/DPS}}
    - \sigma_{\text{DPS},\, y_+\to 0 \rule{0pt}{1.3ex}}
    \ms\right]
\nonumber \\
  & \quad + \left[\ms \sigma_{\text{SPS}\rule{0pt}{1.3ex}}
  - \sigma_{\text{DPS/SPS},\, y_+ \rightarrow 0}
  - \sigma_{\text{SPS/DPS},\, y_- \rightarrow 0}
  + \sigma_{\text{DPS},\,y_\pm\rightarrow 0 \rule{0pt}{1.3ex}} \ms\right]\,,
\end{align}
describing the nested subtraction structure of the full combination. For further details on the different terms, scale setting etc. see in particular section 4.2 of \cite{Diehl:2017kgu} and section 6.5 of \cite{Buffing:2017mqm}. 

%
%
A discussion of the perturbative order at which the ingredients in the SPS as well as DPS cross sections and logarithmic accuracy achievable is given in section 6.6 of \cite{Buffing:2017mqm}. For the colour singlet representation, the ingredients are to a large extent recyclable from resummation for single parton scattering and therefore allow for very high logarithmic accuracy. 

\section{Status of factorisation}\label{sec:factstat}

Essentially all the steps towards a proof of factorisation for TMD DPS producing two uncoloured systems have now been completed \cite{Diehl:2011yj, Diehl:2015bca, Buffing:2017mqm, Vladimirov:2017ksc}. Many of the steps have been achieved by adapting the methodology for the corresponding steps in the proof of factorisation for TMD SPS producing an uncoloured system \cite{Bodwin:1984hc, Collins:1985ue, Collins:1988ig, Collins:2011zzd} (for a brief review of these steps, see \cite{Boer:2017hqr} or \cite{Diehl:2015bca}). 

In derivations of factorisation theorems the most difficult momentum region to treat is the Glauber region -- this region is characterised by the momentum of the particle being mainly transverse (technically, a Glauber momentum $r$ satisfies $|r^+r^-| \ll \tvec{r}^2$). For attachments of Glauber gluons into the two collinear sectors, one cannot apply the so-called Grammer-Yennie approximations \cite{Grammer:1973db, Collins:2011zzd} that one uses for the (central) soft and collinear momenta to separate lines with these momenta into separate functions. A proof of Glauber gluon cancellation for TMD DPS was presented in \cite{Diehl:2015bca}. The argument is based on unitarity, and can be cast into a form where it is rather similar to the corresponding cancellation argument for single scattering found in \cite{Collins:1988ig, Collins:2011zzd}. 

A further important step of the proof is to show that the central soft gluon attachments into the collinear factors can ultimately be disentangled into attachments into the soft Wilson lines given in \eqref{WL-ops}, via nonabelian Ward identities. This step was achieved recently in \cite{Diehl:2018wfy}.

To achieve a description of the TMD cross section in which the soft factor is absorbed into the separate TMDs, it is necessary that the soft factor has the property described in \eqref{S-decomp}. An all-order proof of this property has been put forward in \cite{Vladimirov:2017ksc}, albeit using a different rapidity regulator from that used in \cite{Buffing:2017mqm}. In \cite{Buffing:2017mqm} it was demonstrated that, provided the soft factor can be decomposed in this way, the `splitting' of the soft function into DTMD distributions free from rapidity divergences can be achieved; in particular, the more complicated colour structure compared to SPS does not spoil the factorisation procedure. The systematic combination of SPS and DPS, avoiding any double counting, discussed in the previous section was another important step to obtain a fully consistent DTMD factorisation theorem including both types of scatterings \cite{Diehl:2017kgu}.

At this point the factorisation status for TMD DPS is essentially at the same level as that for TMD SPS. One remaining technical issue relates to Wilson line self-interactions; see section 9 of \cite{Diehl:2017wew} for more information.


\section{Correlations in DPS}\label{sec:corr}
One of the most exciting aspects of DPS is the access it provides to the correlations between two partons bound inside the proton. The study of these correlations actually dates back all the way to the 80s \cite{Mekhfi:1985dv,Mekhfi:1988kj}, but recent years have seen a revival of activity in this area. The two partons can be correlated either kinematically or through their quantum numbers, for recent reviews we refer to \cite{Blok:2017alw,Kasemets:2017vyh} and references therein. Model calculations of DPDFs suggest that such correlations can be large, at least at large values of $x$ \cite{Kasemets:2016nio,Rinaldi:2018zng,Rinaldi:2016mlk,Broniowski:2016trx,Broniowski:2013xba,Rinaldi:2014ddl,Rinaldi:2013vpa,Chang:2012nw}; however, evolution to large scales tends to decrease their importance, especially at low $x$ \cite{Manohar:2012jr,Diehl:2014vaa}. It was recently demonstrated that spin correlations can have a measurable impact on differential distributions in at least one process (same-sign $WW$ production) at the LHC \cite{Cotogno:2018mfv}.

The description of spin correlations in DPS have many parallels and similarities to the spin correlations in TMD physics. The main difference is actually in the physical interpretation, where the correlations between the hadron spin and the spin of a parton is replaced by parton-parton spin correlations. The correlations between the transverse momenta and spin of a parton and/or proton in SPS TMDs are in DTMDs replaced by correlations involving (one or two) parton spins, their two transverse momenta and the transverse distance between them.  The physical difference means that intuition on the size of the correlations built up from SPS can not be applied, but the similarities in the calculations means that large parts of calculations for TMDs can be recycled for DTMDs or DPDFs. 

Colour correlations in DPS do not have a simple analogy in QCD TMD studies; however, certain similarities may be found in recent work on resumming electroweak (EW) logarithms (i.e. logarithms of the hard scale $Q$ over the electroweak scale $M$ when $Q \gg M$), see \cite{Ciafaloni:2001mu, Ciafaloni:2005fm, Bauer:2017isx, Fornal:2018znf,Manohar:2018kfx}. An important difference in resumming electroweak logarithms, compared to QCD, is that the initial state carries $SU(2)$ charge. This means that there can be correlations between the proton EW charge and the charge of the partons probed, which is analogous to correlations between the (colour) charge of two partons inside the proton in DPS. 

\section{Forecast}\label{sec:for}
The recent progress in DPS in general and for DTMD factorisation in particular, have made available the different ingredients necessary for a lot of interesting phenomenology. What has never been done, is to take the theoretical framework, produce all the ingredients and apply it to a particular process such as same-sign W-boson production or double Drell-Yan. This, however, is a rather daunting task if one aims to directly treat all possible correlations, all momentum regions, interferences etc. A more approachable way might be to start the development towards this goal, step by step. It would, for example, be very interesting to see the results of the RGE and rapidity evolution on the DPDs, in particular for the different colour channels. While it is known that the colour correlations decrease with evolution scale, this has never been investigated taking the full evolution equations into account, and simplified studies only explicitly treated the quark non-singlet DPDF \cite{Manohar:2012jr}. A different, interesting and complementary path would be to study the contribution only from the colour singlet DTMDs, and the effect the combination of the regions of $y$ in the DPS cross section has on the transverse momentum distribution of the two vector bosons. On the fixed order computational side, single parton scattering calculations could, with mere changes of colour factors produce important input for DPS. The, by far, largest uncertainty on the transverse momentum dependent DPS cross section comes from the DTMDs themselves. Recent work enables us to make use of as much perturbative information as possible, but there is a lot of potential in model calculations, lattice, evolution studies and extrapolations of the DTMDs into the region of large $z_i$. The latter region is challenging from the modelling perspective, as it requires the treatment of non-perturbative functions depending on three different transverse directions ($z_1$, $z_2$ and $y$). With the LHC ultimately able to reach the integrated luminosity required for detailed studies of the most clean DPS processes, it is envisaged that experimental efforts will be able to put important constraints on the DPDFS, DTMDs and the correlations between two partons inside a proton.

\section*{Acknowledgements}
TK acknowledges support from the Alexander von Humboldt Foundation.

%

\bibliographystyle{jhep}
\bibliography{TMDinDPS}

\end{document}